\documentstyle[12pt]{article}
\normalbaselines
\begin{document}

\begin{center}
{\large \bf
NONLINEAR MAXWELL EQUATIONS}\\[2mm]
G.A. Kotel'nikov\\{\it123182, RRC "Kurchatov Institute", Moscow,
Russia}\\
e-mail: kga@kga.kiae.su\\[3mm]
\end{center}

\begin{abstract}
\smallskip

     A new relativistic invariant version of nonlinear Maxwell 
equations is offerred. Some properties of these equations are 
considered.
\end{abstract}

    The nonlinear equations of theoretical  and  mathematical  physics
attract  the  significant attention because of the specific properties
such, as the absence of the superposition principle, the nonlinear
fields interactions,  existence of the soliton solutions.  The
nonlinear
equations in electrodynamics were offered for the first time  by  Born
\cite{Bor34}, Born and Infeld \cite{Bor34'}, and also Schr\"odinger
\cite{Iva51} outcoming from the
variational principle.  Later Fuschich and Tsifra \cite{Fus85},
Fuschich \cite{Fus92},
Fuschich, Tsifra  and  Bouko \cite{Fus94} have applied to this  goal
the theoretic-algebraic approach.  The purpose of the present work is the
formulation  of
Lorentz and Poincar\'e-invariant equations with the help of the variable
replacement method.

    Let us  introduce the one dimensional Lorentz transformations
\cite{Lan73}

\begin{equation}
x_1'={{x_1 - \beta ct}\over \sqrt {1-\beta ^2}}; \ x_2'=x_2; \
x_3'=x_3; \ t'={{t-\beta x_1/c}\over \sqrt {1-\beta ^2}}
\end{equation}

\noindent Here $x_{1,2,3}= x,y,z$;  $c$ is the speed of light; $t$ is the
time; $\beta = V/c$;
$V$ is the velocity of movement of inertial frame K' relative to K.
\par     One can  see by  calculation  that  the  nonlinear
equations

$$
\Phi_1(I_1,I_2) \nabla .\ {\bf E} = 4\pi \rho; \ {}
\Phi_1(I_1,I_2) (\nabla \times {\bf H} - {1\over c}\partial_t
{\bf E})=+4\pi \rho {\bf v};
$$
\begin{equation}
\Phi_2(I_1,I_2) \nabla .\ {\bf H} = 4\pi \mu; \ {}
\Phi_2(I_1,I_2) (\nabla \times {\bf E} + {1\over c}\partial_t
{\bf H}) = -4\pi \mu {\bf w}
\end{equation}

\noindent are invariant relative to transformations (1), if the values
entering into them  will  be transformed in the known way \cite{Lan73}:

\begin{equation}
E_1' = E_1; \ E_2' = {{E_2 -\beta H_3}\over \sqrt {1-\beta ^2}}; \
E_3' = {{E_3 +\beta H_2}\over \sqrt {1-\beta ^2}};
\end{equation}

\begin{equation}
H_1' = H_1; \ H_2' = {{H_2 +\beta E_3}\over \sqrt {1-\beta ^2}}; \
H_3' = {{H_3 -\beta E_2}\over \sqrt {1-\beta ^2}};
\end{equation}

\begin{equation}
\rho' = \rho {{1 - v_1 V/c^2}\over \sqrt {1-\beta ^2}};
\end{equation}

\begin{equation}
\mu' = \mu  {{1 - w_1 V/c^2}\over \sqrt {1-\beta ^2}};
\end{equation}

\begin{equation}
v_1' = {{v_1 - V}\over {1- v_1 V/c^2}}; \
v_{2,3}' = v_{2,3} {\sqrt {1-\beta^ 2}\over {1-v_1 V/c^2}};
\end{equation}

\begin{equation}
w_1' = {{w_1- V}\over {1- w_1 V/c^2}}; \
w_{2,3}' = w_{2,3} {\sqrt {1-\beta^ 2}\over {1-w_1 V/c^2}}
\end{equation}

\noindent Here $E$, $H$ are the electric and magnetic fields; $\rho$,
$\mu$ are the densities
of electromagnetic  charges;  $v$,  $w$  are  the  charge  velocities;
$\Phi_1$, $\Phi_2$ are
arbitrary functions of  Lorentz  and  Poincar\'e  invariants  of  fields
$I_1=2({\bf E}^2-{\bf H}^2)$, $I_2=({\bf E.H})^2$ \cite{Iva51}.
    In proofing  the invariance of the equations it is necessary to take
into account:  the invariance
of the speed  of light;  the transformation properties of the
electromagnetic fields and charge densities; the law of transformation of
velocities; the invariance of the functions $\Phi_1$ and $\Phi_2$.

    Because of arbitrariness of the functions $\Phi_1$ and $\Phi_2$ system
(2) contains the infinite set of particular realizations of
nonlinear Maxwell equations, among which it is  possible to indicate
the following base versions:
\begin{itemize}
\item  the linear free Maxwell equations \cite{Iva51} with $\rho=\mu=0$;
\item  the linear one-charge Maxwell equations
\cite{Iva51} with $\Phi_1=\Phi_2=1$, \ $\mu=0$;
\item  the linear two-charge  Maxwell  equations  \cite{Str75} with
$\Phi_1=\Phi_2=1$.
\end{itemize}
Let us note some general properties of  these nonlinear  equations induced.
\par The  equations (2) become not only relativistic,  but also conformal
invariant, if  the functions are $\Phi_1 (I_1^ 2/I_2)$,  $\Phi_2 (I_1^
2/I_2)$. The statement
results from the proof of conformal symmetry of linear Maxwell
equations and identical conformal dimension of the values $I_1^2$
and $I_2$.
\par Equations  (2)  become linear in absence of currents and charges and
so contain the classical electrodynamics of the free fields.
\par Generally, the  nonlinearity    is conditioned by currents and
charges.
\par The  equations keep the possibility of electromagnetic field
definition through the two-potentials  $A^a= (\phi,{\bf A})$,
$B^a=(\Phi,{\bf B})$,  $a=0,1,2,3$ \cite{Cab82}

\begin{equation}
{\bf E}= -\nabla \phi- \partial_t {\bf A}/c - \nabla \times {\bf B};
{\bf H}= -\nabla \Phi -\partial_t {\bf B}/c + \nabla \times {\bf A}
\end{equation}

The two-potentials satisfy  the nonlinear D'Alembert equations

\begin{equation}
\Phi_1(I_1,I_2)\Box A^a=4\pi J^a; \ {} \Phi_2(I_1,I_2)\Box B^a=4\pi K^a
\end{equation}

\noindent under condition of the relativistic invariant calibrations
$\partial_a A^a=0$, $\partial_a B^a=0$,
where $\partial_a={\partial}/\partial x^a$, $x^0=ct$,
$x^{1,2,3}=x,y,z$;  $g_{ab}= diag (+,-,-,- )$; $J^0=\rho$,
${\bf J}=\rho {\bf v}/c$; $K^0=\mu$,  ${\bf K}=\mu {\bf w}/c$.
Similarly to the initial equations (2),  the  free
equations (10) automatically become linear.
\par In the important  particular case  of  electrostatic  
charges  in  the one-charge
electrodynamics with ${\bf A}=0$ the scalar potential $\phi$ satisfies
the  nonlinear Laplace-Poisson equation

\begin{equation}
\Phi_1((\nabla \phi)^2) \bigtriangleup \phi = -4\pi \rho (x)
\end{equation}

\noindent Putting here the Fourier-decomposition of the potential
$\phi=(2\pi)^{-3} \int \phi_k exp(i{\bf k.x}) d^3 k$ in case of
electrical charge with the density \ $\rho (x)$
for the component $\phi_k$ we have

\begin{equation}
\phi_k = (4\pi /k^2) \int \rho (x) F_1((\nabla \phi )^2)
exp(-{\bf k.x}) d^3 x = (4\pi /k^2) (\rho F_1)_k
\end{equation}

\noindent Here $(\rho F_1)_k$ means the form-factor characterizing 
the electricity distribution
in the effective charge $Q=\int \rho F_1 d^3 x$, $F_1=1/\Phi_1$. The
form-factor can differ from
unit.  This will mean the availability of corrections to the Coulomb 
field of a charge.
\par Let  us  write  the equations (2) in some other form.   We divide
their right parts into the functions $\Phi_1$ and $\Phi_2$,  designate
$1/\Phi_1=F_1$, $1/\Phi_2=F_2$
and instead of $\rho(x)$, $\mu(x)$, ${\bf J}=\rho(x) {\bf v}/c$,
${\bf K}=\mu(x) {\bf w}/c$ we take the new values

\begin{equation}
\rho \to \rho (x) F_1; \ {} \mu \to \mu (x) F_2; \ {}
{\bf J} \to \rho(x) F_1 {\bf v}/c; \ {} {\bf K} \to \mu(x) F_2 {\bf w}/c
\end{equation}

\noindent We will refer  the densities of  charges and
currents $\rho$, $\mu$, ${\bf J}$\ and ${\bf K}$ to as the initial ones,
and the values   corresponding them do to as  the effective  ones.
Then it is possible to say that the nonlinear microscopic
equations of electrodynamics are the equations  which
contain the effective values of charge densities and
current densities  instead of the initial ones \cite{Kot94}

$$
\nabla .{\bf E}=4\pi F_1(I_1,I_2) \rho ; \ {}
\nabla \times {\bf H} -{1\over c}{\partial_t} {\bf E}=
4\pi F_1(I_1,I_2) \rho {{\bf v}\over c};
$$
\begin{equation}
\nabla .{\bf H}=4\pi F_2(I_1,I_2) \mu; \ {}
\nabla \times {\bf E} +{1\over c}{\partial_t} {\bf H}=
-4\pi F_2(I_1,I_2) \mu {{\bf w}\over c}
\end{equation}

\begin{equation}
\Box A^a = 4\pi F_1 {\bf J}^a; \ {} \Box B^a = 4\pi F_2 {\bf K}^a
\end{equation}

\noindent These equations realize the principle of self-action: the 
initial charges generate electromagnetic fields which in its turn 
influence on the initial
charges,  their densities and sizes up to  reaching  the  equilibrium
state with the generating fields. So, in the nonlinear versions of the 
Maxwell and D'Alembert equations (14), (15) the electromagnetic charges

\begin{equation}
Q=\int \rho(x) F_1(I_1,I_2) d^3 x; \ {}
P=\int \mu(x)  F_2(I_1,I_2) d^3 x
\end{equation}

\noindent receive at least partly the field nature.  This property  of
a charge is
absent  in the linear  electrodynamics.  The  effective  charges $Q$ 
and $P$ keep the property of \ Lorentz-invariance owing to the invariance 
of functions $F_1$ and $F_2$, and are integrals of movement due to
the existance of the continuity  equations

\begin{equation}
\partial_t(\rho F_1) + c\nabla .(F_1 {\bf J})=0; \ {}
\partial_t(\mu  F_2) + c\nabla .(F_2 {\bf K})=0
\end{equation}

\noindent  It followes  from  the equations  (17)  that the initial 
charges are not
conserved.  For example, we have  for the electric  charge  $q=\int
\rho d^3 x$ 

\begin{equation}
\partial_t q = -\oint \rho {\bf v} ds - \int (\partial_t F_1 +
{\bf v}.\nabla F_1)(\rho /F_1) d^3 x
\end{equation}

\noindent Here $ds$ is the element of the area surrounding the volume  
element $d^3 x$  as
usually. The  change  of  the  charge $q$ is conditioned not only by the
density of current ${\bf j}=\rho {\bf v}$,  but  by the change  of
the  field  invariant $F_1(I_1,I_2)$  in time and space.
\par  As far as the nonlinear Maxwell equations satisfy  the  requirement
of  relativistic  invariance,  they have the potential interest to
physics.  In addition to the known general theoretical questions of
electrical  charge  stability and nature of its mass \cite{Iva51},
it is possible
to point out also the field nature of a charge and the necessity of
experimental verification of the Coulomb law at short distances.
     The existance of equations (14) prompt us also to induce  the
relativistic invariant action integral in case of the one charge
electrodynamics  in more general form

\begin{equation}
S=-mc\int \Psi_1(I_1,I_2) ds - {1\over c}\int \Psi_2(I_1,I_2) A_a J^a
d^4 x - {1\over 16\pi c}\int \Psi_3(I_1,I_2) I_1 d^4 x
\end{equation}

\noindent Here as usually,  $m$ is the rest mass of a particle, $ds$
is the element of
the interval, $A_a=(\phi ,-{\bf A})$, $d^4x=cdtdxdydz$ \cite{Lan73},
$\Psi_1$, $\Psi_2$, $\Psi_3$ are
the functions of relativistic invariants $I_1$ and $I_2$.

According to (19) we can  indicate  the six
versions of Maxwell
electrodynamics with the invariant  speed of light: 
\begin{itemize}
\item the classical linear electrodynamics with $\Psi_1=\Psi_2
=\Psi_3=1$ \cite{Iva51}, \cite{Lan73};
\item  the linear electrodynamics  with $\Psi_1 \neq 1$,
$\Psi_2=\Psi_3=1$;
\item  the nonlinear electrodynamics of the first type 
with $\Psi_3 \neq 1$, $\Psi_1=\Psi_2=1$;
\item  the nonlinear electrodynamics of the second type 
with $\Psi_2 \neq 1$, $\Psi_1=\Psi_3=1$;
\item the nonlinear electrodynamics of the third type 
with $\Psi_2 \neq 1, \Psi_3 \neq 1, \Psi_1=1$;
\item the nonlinear electrodynamics of the fourth type 
with all functions $\Psi \neq 1$.
\end{itemize}
In particular, the Born model   with
$\Psi_3=4E_0^2 [1- (1-I_1/2E_0^2)^{1/2}]/I_1$ \cite{Bor34},
the Born-Infeld model with
$\Psi_3=4E_0^2 [1-(1-I_1/2E_0^2-I_2/4E_0^4)^{1/2}]/I_1$ \cite{Bor34'},
the Schr\"odinger   model  with  $\Psi_3=2E_0^2 \ ln(1+I_1/2E_0^2)$
\cite{Iva51} belong to the nonlinear version of the first type. (Here
$E_0$ is the maximum field \cite{Iva51}).
\par This work belongs to the nonlinear  version of the second type,
as far as the variation of the
integral (19) with  the invariable value of $\Psi_2 J^a$
leads to the equations (14). For example,  within  framework of this
version  the  nonlinear
Laplace-Poisson equation (11) may be written as follows:

\begin{equation}
[1+ \alpha ({\partial \phi\over \partial r})^2][({1\over r^2})
{\partial \over \partial r}(r^2{\partial \phi \over \partial r})]=
\cases{-{2q\over r^2}({a\over \sqrt \pi} e^{-a^2r^2})&if $\rho=\rho_1$;
\cr\noalign{\vskip 4pt} -{2q\over r^2}({a\over \pi}{sin {} ar\over ar})&if 
$\rho=\rho_2$ \cr}
\end{equation}
\smallskip

\begin{sloppypar}
\noindent Here we put that $\Phi_1=[1+\alpha I_1]=
[1+\alpha {\bf E}^2]_{H=0}=
[1+\alpha (\partial \phi/\partial r)^2]$; $\rho_1 = (q/2\pi r^2)(a/\sqrt
\pi)exp(-a^2 r^2)$; $\rho_2 =(q/2\pi r^2)(a/
\pi)(sin ar/ar)$, $q$ is the electrical charge, $r=(x^2+y^2+z^2)^{1/2}$, 
$\alpha=k/a$, $k$ is the proportionality
coefficient, $a$  is the parameter with inverse length dimension (1/cm).
Tending
$a \to \infty$ ($\alpha \to 0$), we have 
the linear Laplace-Poisson equation
\begin{equation}
{1\over r^2}{\partial \over \partial r}(r^2{\partial \phi \over
\partial r})=-{2q\over r^2}\delta (r), 
\end{equation}
where the solution of the equation has the known form $\phi = {q/r}$. 
One can  see, that the linear theory is the theory with  point-charges.
The nonlinear theory is the theory with the density of charges 
distributed over space.
Densities of charges $\rho_1$ and $\rho_2$
correspond to the various  physical models of charges. 
\end{sloppypar}
\par The others versions were not investigated. They can be accompanied
by the dependence of the effective mass on the electromagnetic field.
\par In addition to these six versions, it is possible to formulate 
the nonlinear  electrodynamics with the noninvariant  speed of
light. The model of this type was proposed by Fushchich \cite{Fus92}.


\begin{thebibliography}{99}
\bibitem{Bor34} M. Born. Proc. Roy. Soc. A, 1934, v. 143, N 848, p.
410-437.
\bibitem{Bor34'} M. Born, L. Infeld. Proc. Roy. Soc. A, 1934, v. 144,
N 850, p. 425-451.
\bibitem{Iva51} D. Ivanenko, A. Sokolov. Classical Theory of Field.
Œoscow-Leningrad: Gostexizdat, 1951, p. 199.
\bibitem{Fus85} W.I.  Fuschich,  I.M.  Tsifra. Teor. Mat. Fizika
(Russia), 1985, v. 64, N 1, p. 41-50.
\bibitem{Fus92} W.I.  Fuschich. Dok. Acad. Nauk (Ukraina), 1992, N 4,
p. 24-27.
\bibitem{Fus94} W. Fuschych, I. Tsyfra, V. Bouko. Nonlinear Math.
Phys. (Ukraina), 1994, v.1, N 2, p. 210-221.
\bibitem{Lan73} L.D. Landau, I.M. Lifshits. Theory of Field. Œoscow:
Nauka, 1973, p. 89, 102.
\bibitem{Str75} V.I. Strazhev, L.M. Tomilchik. Electrodynamics with
Magnetic Charge. Minsk: Nauka i Technika, 1975, p. 18, 41.
\bibitem{Cab82} N. Cabibo,  E. Ferrari.  Nuovo Cimento,  1982,  v.
23,  N. 6, p. 1147-1154.
\bibitem{Kot94} G.A. Kotel'nikov. Izv. VUZov (Russia), 1995, N 2, 
p. 116-119.
\end{thebibliography}
\end{document}